# Reusability Challenges of Scientific Workflows: A Case Study for Galaxy

Khairul Alam
*Department of Computer Science*
*University of Saskatchewan*
Saskatoon, Canada
kha060@usask.ca

Banani Roy
*Department of Computer Science*
*University of Saskatchewan*
Saskatoon, Canada
banani.roy@usask.ca

Alexander Serebrenik
*Department of Mathematics and Computer Science*
*Eindhoven University of Technology*
Eindhoven, Netherlands
a.serebrenik@tue.nl

*Abstract*—Scientific workflow has become essential in software engineering because it provides a structured approach to designing, executing, and analyzing scientific experiments. Software developers and researchers have developed hundreds of scientific workflow management systems so scientists in various domains can benefit from them by automating repetitive tasks, enhancing collaboration, and ensuring the reproducibility of their results. However, even for expert users, workflow creation is a complex task due to the dramatic growth of tools and data heterogeneity. Thus, scientists attempt to reuse existing workflows shared in workflow repositories. Unfortunately, several challenges prevent scientists from reusing those workflows. In this study, we thus first attempted to identify those reusability challenges. We also offered an action list and evidence-based guidelines to promote the reusability of scientific workflows. Our intensive manual investigation examined the reusability of existing workflows and exposed several challenges. The challenges preventing reusability include tool upgrading, tool support unavailability, design flaws, incomplete workflows, failure to load a workflow, etc. Such challenges and our action list offered guidelines to future workflow composers to create better workflows with enhanced reusability. In the future, we plan to develop a recommender system using reusable workflows that can assist scientists in creating effective and error-free workflows.

## I. INTRODUCTION

Scientific workflows (SWs) serve as a valuable resource for researchers in various fields, enabling them to enhance their research and development procedures and attain outcomes that are both more reliable and reproducible. SWs are adept at the streamlined analysis of vast datasets, a necessity for numerous research endeavors addressing the most critical and intricate global issues, including tasks such as plant phenotyping and genotyping for food security, forecasting water quality, and conducting population-based cancer investigations [2], [5]. Furthermore, with the growing volume, diversity, and speed of data generation, there is a rising trend in conducting extensive data experiments using Scientific Workflow Management Systems (SWfMSs). These systems are employed to investigate data, strategize the execution of experiments, and visualize the findings. In a SWfMS, the computational data analysis involves a series of stages, including data preprocessing, quality control, quantification, and statistical analysis, all aimed at transforming raw data into meaningful scientific outcomes. Thousands of software tools have been created to carry out this computational analysis and visualization, catering to a wide range of SWfMSs. To illustrate, consider the Galaxy [1] SWfMS, which has over 8400 accessible and validated software tools [6]. The utilization of this vast array of tools poses a significant challenge for workflow practitioners when constructing SWs. Research studies [2], [7], [8], have pointed out that despite decades of effort, current SWfMSs have not adequately addressed the requirements of the global scientific community for swiftly creating and adapting intricate workflows. Up to this point, determining the precise sequence of modules and connections necessary to achieve a desired outcome remains a daunting task without an in-depth understanding of the underlying computational components. Moreover, identifying which software tools and combinations are practical or optimal in practical scenarios often proves to be challenging. Consequently, researchers encounter difficulties when selecting suitable and efficient data analysis tools for a particular experimental design, as highlighted by Palmblad et al. [41]. Consequently, researchers remain unassisted in creating relevant analyses from these resources that remain largely underutilized [8].

Analyzing scientific data is one of the most challenging issues scientists face today. As analyses get more complex and extensive, interdisciplinary groups need to work together, and knowledge sharing becomes essential to support effective data exploration [19]. So, researchers developed several community frameworks for analyzing workflows. As a result, public repositories are gaining importance as a means to share, find, and reuse such workflows [16]. Moreover, open science has emerged as a framework for improving the quality of scientific analysis. Transparent, accessible knowledge-sharing and collaborative networks are essential components of open science. SW communities are also tending toward open science, and as a result, they make many workflows available to the community in different repositories. Scientists share their workflows with these repositories so that other users

This research is supported by the Natural Sciences and Engineering Research Council of Canada (NSERC) Discovery, Collaborative Research and Training Experience (CREATE), and graduate program on Software Analytics Research (SOAR) grants.

can reuse them to solve their problems. Reusing workflows allow scientists to do new experiment faster, as the workflows capture valuable expertise from other scientists. Unfortunately, these workflows are not always reusable by other users. There are several [3], [9], [16], [40], [48], [49] studies using available workflow repositories to reuse existing workflows in different ways, but there is a marked lack of research investigating (1) *the challenges of reusability of scientific workflow* and (2) *the approaches to overcome such challenges*.

SWs and SWfMSs facilitate the process of conducting scientific experiments and analyzing results. They empower researchers to automate mundane tasks, effectively handle intricate data, and foster productive collaboration among peers. SWs also enhance the reproducibility and transparency of research by providing a clear record of the steps taken to arrive at a particular result. In addition, SWfMSs further improve SWs by offering tools for version control, error handling, and data visualization. Furthermore, SWfMSs enhance SWs by equipping them with version control, error management, and data visualization capabilities. In essence, these technologies contribute significantly to simplifying the scientific process, resulting in more efficient and rigorous research.

Among SWfMSs, Galaxy [1], a web-based, open-source framework, has gained widespread adoption for large-scale data analysis, boasting over 169 research server installations worldwide [21]. It provides facilities for managing and ensuring FAIR data (Findable, Accessible, Interoperable, and Reusable) analysis. It enables researchers to use thousands of software tools from various domains and provides an interactive environment for researchers. It offers powerful and practical solutions for analyzing scientific data by providing access to extensive hardware, tools, and data that can be adopted with relatively minimal training [66]. Galaxy boasts a thriving community, actively maintaining servers, tool integration, hosting workshops, and addressing queries [17]. The community is rapidly growing [67], and a dedicated help forum facilitates Q&A [18]. Moreover, it is the most successful SWfMS [20] and intuitive to use. Its status as the most successful SWfMS is attributed to user-friendliness, accessibility, modularity, reproducibility, strong community support, and seamless tool integration. Thus, for conducting this research work, we mainly focused on Galaxy-based scientific workflows.

In this paper, we followed three sequential steps for each scientific workflow. First, we attempted to understand the purpose of the workflow by reading the name and annotation. Then we imported the relevant workflow on our workflow list (you need to register to Galaxy for importing) and tried to reuse it. Finally, we recorded our findings on the challenges of reusing a SW, and in case of failure to reuse, we investigated why it cannot be reused. We also recorded the actions necessary to make a SW reusable. We answered three research questions and thus made three contributions as follows:

**(RQ1) What are the challenges to reuse a scientific workflow? How can reusability challenges be measured?** We conducted an extensive manual analysis and investigated the reusability challenges of SWs related to the Galaxy. We classified the reusability status of SWs into *Reusable* and *Nonreusable*. We further sub-categorized the reusable workflows into *Reusable without modification, Easily reusable, Moderately difficult to reuse and Difficult to reuse*.

**(RQ2) What percentage of scientific workflows can be reused or not with or without modifications?** We conducted an extensive statistical analysis to assess the percentage of workflows that can be reused, either with or without slight or substantial adjustments to the original workflows.

**(RQ3) How can we overcome the reusability challenges?** While we tried to reuse SWs, we encountered several challenges. We had to perform several minor or major changes to make the workflows reusable. We meticulously documented a series of actions taken in this regard.

## II. BACKGROUND

This section introduced Scientific Workflow (SW), Scientific Workflow Management Systems (SWfMSs), and Scientific Workflow Repositories (SWRs). First, we defined SW, SWfMSs, and SWRs. Then, we discussed the relationship among Software Engineering (SE), SWs, and SWfMSs.

### A. Scientific Workflow (SW)

A scientific workflow (SW) is a structured and automated approach for designing, executing, and analyzing scientific experiments or computational processes. It involves using software tools and frameworks to create a sequence of interconnected tasks or steps that define the entire scientific process, from data input to analysis and result generation. SWs often leverage concepts and techniques such as modularization, abstraction, encapsulation, and separation of concerns. They involve using programming languages, scripting, and workflow management systems to define, execute, and monitor the progress of the workflow. Additionally, SWs may integrate with external tools, libraries, databases, and data sources to handle data processing, analysis, visualization, and other scientific tasks.

Figure 1 shows an example of a scientific workflow obtained from WorkflowHub. This workflow assess genome quality by generating some statistics and determining if expected genes are present; align contigs to a reference genome. The inputs of the workflow are polished assembly and reference_genome.fasta. Here, two software tools are *Busco, Quast* used. The outputs contain Busco table of genes, Quast HTML report, etc. There are several parameters for the tools, but we are not discussing everything in detail due to space constraints.

The workflow steps are known as *Computational Modules*. The computational modules are responsible for data manipulation and processing. A workflow module starts its execution when the required datasets are available.

### B. Scientific Workflow Management Systems (SWfMSs)

Scientific workflow management systems (SWfMSs) are specialized software tools or platforms designed to facilitate the creation, execution, monitoring, and management of SWs. These systems provide a comprehensive environment for scientists, researchers, and engineers to design, deploy, and analyze complex scientific experiments or computational processes. SWfMSs play a critical role in software engineering by providing a unified and streamlined approach to designing, executing, and managing complex scientific processes. Some of the most used SWfMSs are Galaxy [1], Taverna [34], Wings [10], Vistrails [11], Pegasus [12], Kepler [13], Askalon [37] and Knime [14]. Among them, Galaxy is a web-based, community-driven, easily accessible, and open-source platform for scientific data analysis. It generates concrete tasks for each activity, puts the tasks in a queue to be submitted, and monitors the task status.

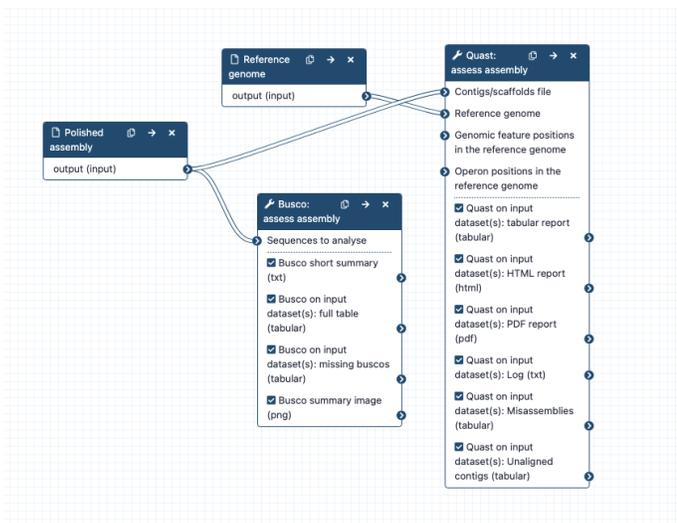

Fig. 1. An example of a scientific workflow from WorkflowHub
Title: Assess genome quality

### C. Scientific Workflow Repositories

Scientific workflow repositories are online platforms that allow researchers to store, access, and share scientific workflows. These repositories typically contain an extensive collection of workflows created by researchers from different domains and are designed to facilitate collaboration and knowledge sharing in the scientific community. A scientific workflow repository typically provides features such as: *search, download, metadata, upload and share, and version control*

Examples of scientific workflow repositories include myExperiment [38], WorkflowHub [22], Galaxy [26], Taverna Workflow Library [68], and Bioinformatics Workflow Repository (Biowep) [69]. Some other popular workflow repositories are [15], [27], [29]–[33], [54]. These repositories are valuable resources for researchers as they can avoid duplication of efforts, speed up their analysis pipelines, and improve the reproducibility of their research.

### D. Software Engineering, Scientific Workflows and SWfMSs

Software engineering provides the foundation and practices for the development and application of scientific workflows. SWs represent scientific experiments or processes, while SWfMSs provide the necessary infrastructure and tools to create, execute, and manage these workflows efficiently. The relationship among these three domains is symbiotic, with each one contributing to the effectiveness and efficiency of scientific research and experimentation. We can consider the SWfMSs like Galaxy, Taverna, LONI pipeline, or Generation as an integrated software framework for performing scientific data analysis.

## III. REUSABILITY

Reusability in SWs refers to the ability to reuse or repurpose existing workflows for different experiments or scientific tasks. It involves designing modular and flexible workflows so that they can be easily adapted and reused by different users or contexts. Reusability offers several benefits (e.g., reproducibility, collaboration and knowledge sharing, time and effort Savings, best practices and optimization, and standardization and interoperability) in scientific research.

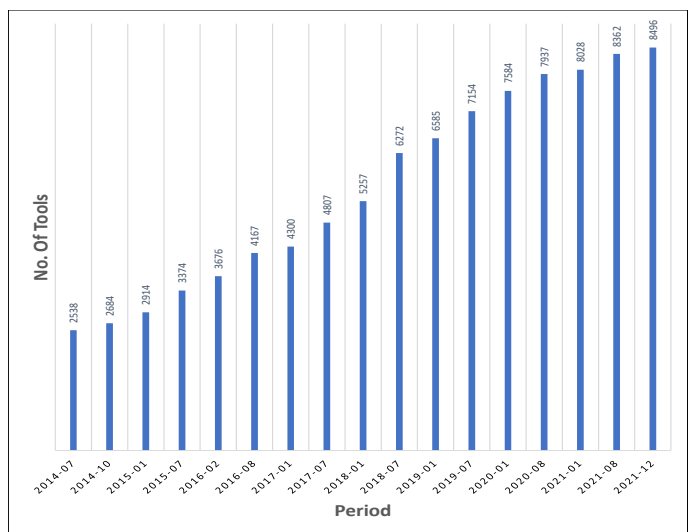

Fig. 2. Galaxy SWfMS tools statistics over period

There are thousands of tools in SWfMSs for performing several computational and visualization tasks, which are increasing rapidly. For example, consider the Galaxy SWfMS tools statistics shown in Figure 2. It is almost impossible for any workflow designer to remember these vast amounts of tool functionalities. In addition, finding which tools and combinations are optimal in workflow construction is often problematic. Scientists can reuse the shared workflows available at different repositories [15], [26], [27], [29], [33], [54] to mitigate these challenges. But reusing existing workflows has several challenges, which motivated us to do this research.

## IV. MOTIVATING EXAMPLES

Scientific workflows published at scientific workflow repositories might not always be reusable. For example, let us consider a fragment of a scientific workflow in Figure 3. The workflow name is *RNA-seq analysis from SRA PE data (hg38, genecode.v24)*. Here, the workflow designer tried to do RNA sequence analysis using human genome SRA pair-end data. Any user can check the workflow using the Galaxy workflow repository [26] by searching the workflow using the name.

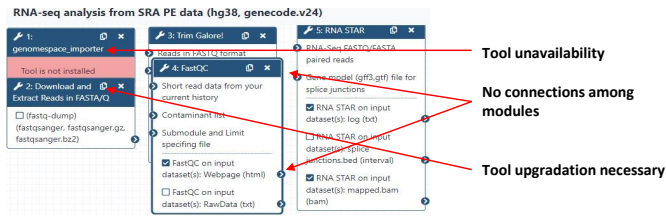

Fig. 3. A segment of a SW (**Reusability status: Nonreusable**)

We attempted to reuse the workflow, but we faced several challenges in doing so. Firstly, *genomespace-importer* tool support is not available in Galaxy as this tool is broken [42]. Then, in a workflow, the tools are connected and perform sequential operations, but there has been no tool connection for this workflow. All the tools used are isolated. And finally, one tool needs up-gradation as indicated in Figure 3. The tool up-gradation can be done easily, but the other challenges are not resolvable.

Let us consider another example, as shown in Figure 4. The name of the workflow is *Creating a bed file of significantly different genes*. In this workflow, the workflow designer performed at first a filter operation using a condition *(c14=='yes')*.

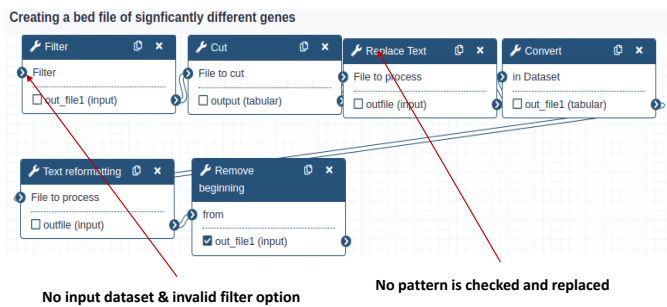

Fig. 4. An example of a SW (**Reusability status: Nonreusable**)

By exploring genome formats [43], we noticed that there are three formats *(BED detail format has 14 columns, ENCODE gappedPeak: Gapped Peaks (or Regions) format has 15 columns, .2bit format has 16 columns)* which have 14 or more columns, and none of these formats contain value *yes* in column 14. Also, the workflow designer did not reveal anything about the input dataset. Another issue with the workflow is that it used a tool *Replace Text* to replace text in a specific column using a pattern. But the workflow designer did not provide any pattern to find and replace. These reasons make the workflow nonreusable.

Let us consider a reusable workflow example shown in Figure 5. The workflow name is *Exploring Iris dataset with statistics and scatterplots*. The workflow has the annotation (*Exploring the Iris dataset and visualizing features...*) and tagging information (*iris, Groups, Scatterplot*).

At each step of action, a proper explanation is given for this workflow. Like where the data can be obtained and how to preprocess the data by converting format and removing the header, an intuitive data analysis process (how many species, count of species, and differentiating the species). Finally, the owner visualized Iris dataset features with two-dimensional

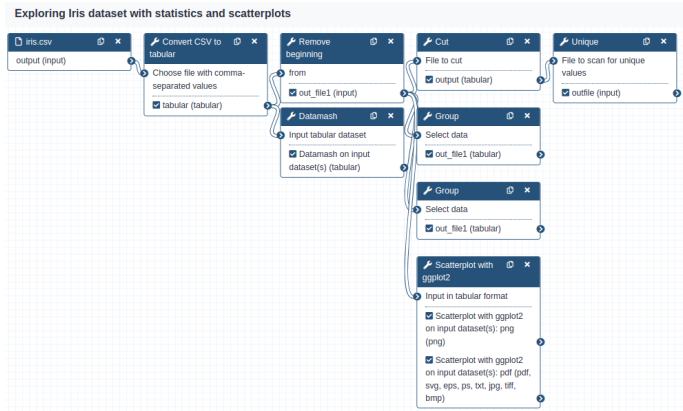

Fig. 5. An example of a scientific workflow (**Reusability status: Reusable**)

scatterplots. All activities are nicely explained in stepwise annotations.

*Easily reusable, Moderately difficult to reuse and Difficult to reuse* workflow examples are demonstrated in Section VI.

## V. STUDY METHODOLOGY

Figure 6 shows the schematic diagram of our conducted exploratory study. We, at first, selected statistically significant (95% confidence level with 5% margin of error) Galaxy SWfMS-related scientific workflows from available repositories [26], [29], [33] and then tried to reuse them. We could reuse a percentage of workflows without facing any issues, but there were many workflows where we needed to perform a list of actions to make them reusable. And we got some workflows that are nonreusable. The following sub-sections discussed different steps of our methodology.

### A. Dataset Preparation

Figure 7 depicts the data collection steps. We collected Galaxy SWfMS-related SWs from the available repositories [26], [29], [33]. These repositories contain 1503 scientific workflows (Feb 2023).

At first, we discarded such workflows that do not have a proper name or annotation. The workflow's name indicates the purpose of the workflow, and the annotation denotes the nature of activities occurring in them. By reading the name, any workflow designer should be able to understand what the workflow is doing. In addition, the annotation helps the workflow designer describe the workflow's particular functionality to the community so that they can easily understand the tasks to reuse. The annotation also helps to create a new workflow by assisting users in selecting suggestive components [44]. Our manual investigation found 17.9% of workflows that do not have proper names/annotations. A sample of improper names are *Unnamed workflow, Unnamed history, Test Workflow, Galaxy Test, Assignment, username, Workshop*, and in most of the cases for these workflows, the annotations are missing. Naming/annotating is a qualitative measure, so we conducted an agreement analysis with our findings with another software researcher *(Ph.D. student and have a fundamental knowledge of scientific workflow and SWfMSs)*. The researcher took (95% confidence level with 5% intervals) workflows name/annotation information from the shared workflow name/annotation document, and we independently marked each name/annotation as proper or improper. We finally measured the agreement using Cohen's Kappa [28]. The value of $\kappa$ was 0.86, meaning the agreement's strength is near perfect.

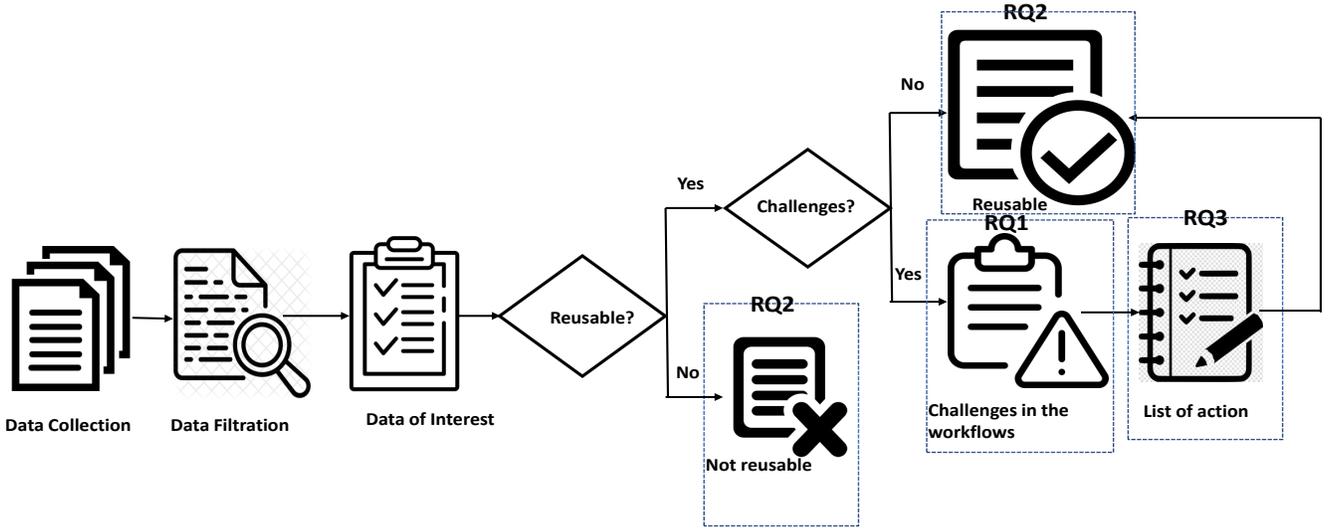

Fig. 6. Schematic diagram of our exploratory study

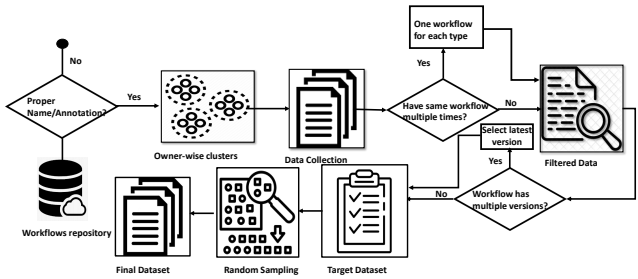

Fig. 7. Selection of dataset for study

Then we made owner-wise clusters using the cluster sampling [45] method. If multiple clusters had the same workflow multiple times, we picked one workflow and discarded the others. Using this approach, we removed 11.96% of workflows. Later, we checked if a workflow had multiple versions (for each cluster); we took the updated one and discarded the others. This way, we eliminated 2.96% of workflows. Finally, from the remaining workflows, we randomly selected 307 workflows for our analysis and answered our research questions.

### B. Data analysis

We manually analyzed 307 workflows. For each workflow, we attempted to understand the purpose of the workflow (the problem it is solving) and the usage of the tools. In addition, we checked the parameter values of tools wherever relevant. Then we tried to reuse every workflow. We often faced several reusability challenges, so we needed to perform trivial, minor, and major edits on the workflows to overcome the challenges. The Study Findings Section VI describes all the things.

In order to identify the challenges the users may face, along with reusability checking, we also checked the community forum [18]. The forum has thousands of issues from users using Galaxy. The challenges include tools not working, different results on the same workflow, complexities in providing tools parameters, missing functions, long-running queries, etc. The users also ask for help constructing workflow or using any particular tool in the forum. The community members responded to most of the questions and provided different solutions for each question. We considered the questions while identifying the challenges and checked the answer to make the action list to resolve the challenges.

## VI. STUDY FINDINGS

By exploring the 307 workflows, we identified several reusability challenges, and to overcome these challenges, we performed several activities. Based on the effort level, we classified the reusability status. We asked three research questions in this study and answered them carefully with the help of our data analysis findings.

### A. RQ1: What are the challenges to reuse a scientific workflow? How can reusability challenges be measured?

We divideed **RQ1** into two parts **RQ1(a) and RQ1(b)** respectively and answered them separately. Table I enlists the challenges we faced in reusing a workflow, and Figure 8 shows our measuring approaches to reusability challenges.

**Answering RQ1(a): What are the challenges to reuse a scientific workflow?** We attempted to reuse our sampled workflows. However, we faced several challenges, and sometimes, we got many workflows that were nonreusable due to several non-trivial reasons. Table I shows the challenges we faced. A workflow may experience multiple challenges. We are discussing each challenge briefly as follows.

**Obsolete tools in the workflows:** There are many obsolete tools in workflows. Tool supports are unavailable for these tools (i.e., Galaxy removed them from its tool repository). There are several reasons for tool removal. The tool may be replaced by a more robust alternative tool; sometimes, the tool may have several bugs, or the same activities can be performed alternatively. By checking each workflow manually, we identified that 12.38% of workflows have one or more deprecated tools. These obsolete tools hinder reusability.

**Tools Update/Upgrade:** Software tools update/upgrade is a very common thing, and workflows tools are not apart from it. Bug fixing, resolving security issues, adding new features, and performance improvement lead to the tools update/upgrade. By analyzing workflows, we found that 39.09% of workflows need tools update/upgrade to make the workflows reusable.

**Workflows cannot be presented as a DAG:** In a SW, the tools are connected sequentially, and the output of one tool can be the input of one/more tools. In most cases, the workflow should be presented as a directed acyclic graph. But we found 10.10% of workflows where the tools are not connected, which eventually violates SWs characteristics.

TABLE I
REUSABILITY CHALLENGES OF SCIENTIFIC WORKFLOWS

| Sl No. | Reusability Challenges |
|---|---|
| 1 | Obsolete tools in the workflows (12.38%) |
| 2 | Tools Update/Upgrade (39.09%) |
| 3 | Workflows cannot be presented as a DAG (10.10%) |
| 4 | Incomplete & Empty workflows (0.77%) |
| 5 | Workflows loading failure (3.91%) |
| 6 | Incompatible tools connection (2%) |
| 7 | Invalid operations in the workflows (1.54%) |
| 8 | Missing/Incorrect values for tool parameters (0.77%) |
| 9 | Multiple independent workflows in a workflow (3.58%) |
| 10 | No information about the input dataset (1.54%) |
| 11 | Misleading workflows (0.77%) |

**Workflows loading failure:** We identified 3.91% of workflows that returned an internal server error (failed to load a workflow in SWfMS). These workflows are nonreusable. Actually, we were unable to check the tools and connections they contained.

**Incompatible tools connection**: In workflows, the tools are connected sequentially, and the output of one tool is the input of others. So the data type of the output tool and associated input tools need to be compatible. But by analyzing, we found that 2% of workflows contained incompatible tool connections.

**Multiple independent workflows in a workflow**: A workflow can contain multiple sub-workflows, but we noticed that some workflows have multiple independent workflows (in a workflow, there have multiple workflows) which are not connected at all. These workflows are doing different operations. We found 3.58% of such workflows in the observed sample. This also increases the challenges of reusability.

There are also several other challenges of reusability, i.e., there are workflows that are doing invalid operations (e.g., cutting a column that is not available in the dataset, and filtering with irrelevant data). We also find some workflows where no information is available about the input dataset; also, it is not understandable by reading the name or annotation they provided. A few workflows do not perform the tasks as it is mentioned in the workflow name. We identified these workflows as misleading workflows. We also got several workflows where the parameter values are missing (Replace Tool is used, but no value is given in finding and replacing patterns) or incorrect values (Filtering is done by a value not compatible with the dataset column). We identified that 5.54% of workflows have these challenges by analyzing our sample workflows.

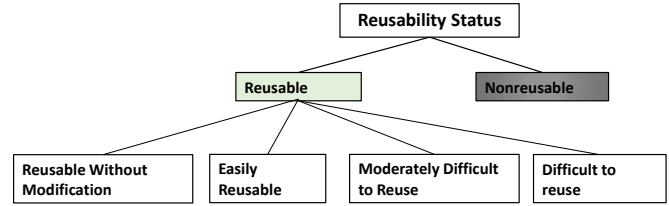

Fig. 8. Classification of reusability status

**RQ1(b): How can reusability challenges be measured?:** We divided the reusability status (of workflows) using two different dimensions. These are reusable and nonreusable. Based on the effort we needed to make a workflow reusable, we further classified the reusable workflows into four sub-categories. Figure 8 shows our sub-classification of the reusable category. In the following section, we discussed the reusability status in detail.

**Reusable:** A significant percentage of workflows can be reused without any modification. But there are many cases where workflows require some modifications. We, therefore, classified the reusable status into four more sub-categories based on the complexity of the modifications and the time required to reuse the workflow. Figure 8 shows our sub-classification of the reusable category.

- *Reusable without modification (RWM):* The given workflow is complete, stand-alone, and requires no explicit action to reuse. The workflow in Figure 5 showed such an example. Any user can use this workflow to distinguish the iris species through visualization.
- *Easily reusable (ER):* These workflows can be reused by performing some modifications that are comparatively less complex (*adding a few connections, importing known input data, tools update/upgrade*) and less time-consuming. The activities we needed to perform here were to update/upgrade one or more tools, adding a connection between tools, and sometimes change the label of the steps. Figure 9 shows such an example. The name of the workflow is *QualStats-Boxplot-Distribution*. This workflow creates a quality statistics report for FASTQ data and shows the result in boxplot and distribution chat. We needed to import the *FASTQ* dataset as input and connect it to the *Compute quality statistics* module.
- *Moderately difficult to reuse (MDR):* These workflows require much modification and are comparatively more complex (*Workflow has few errors, requires manual debugging to find issue-creating tools, finding out alternative tools for deprecated tools*) and time-consuming.

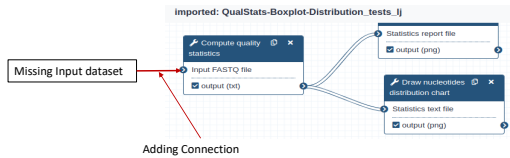

Fig. 9. An example of a SW (**Reusability status: Easily reusable**)

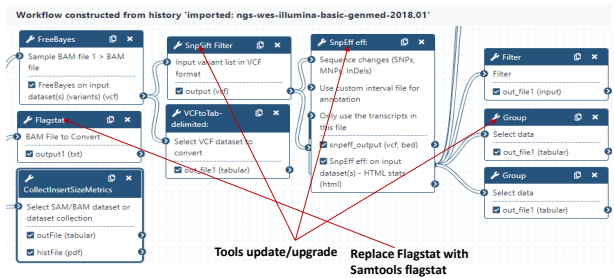

Fig. 10. A segment of a SW (**Reusability status: Moderately difficult to reuse**)

For example, consider a segment of workflow shown in Figure 10. The workflow name is *Workflow constructed from history 'imported: ngs-wes-illumina-basic-genmed-2018.01.'* . While trying to reuse this workflow, we faced errors, then we debugged the workflow and found that the *flagstat* tool is the reason. We replaced this tool with an alternative tool named *Samtools flagstat* by keeping the same outcome. We also needed to update *SnpSift Filter, SnpEff eff, Group* tools.

- *Difficult to reuse (DR):* These workflows are complex and require a lot of modifications (*Workflow has several errors, lots of module connections, failure of tool up-gradation, lots of debugging necessary*), and are very time-consuming. Consider a segment of workflow shown in Figure 11. The workflow name is *workflow constructed from history 'SNP for Trio'*. We needed to update/upgrade *VCFfilter, SnpEff eff, Group, Select first* tools. After that, we faced issues with saving the workflow. Then we debugged the workflow and found that *VCFfilter, SnpEff eff* tools up-gradation failed. Finally, we removed these tools from the workflow with connected modules and later imported the latest versions of these tools and created the associated connections.

**Nonreusable (NR):** We could not reuse these workflows even after several minor and major modifications. We, two authors (more than five years experienced with SWs), took part in the manual analysis. At first, one author marked the workflows as nonreusable. Then, another author randomly took 22% of workflows from the list and tried to reuse them. In 90% cases, we got a similar result (for the rest of the 10%, the disagreement occurred because another author considered that a part of these workflows could be reused). Figure 4 shows an example of a nonreusable workflow.

### B. Answering RQ2

**(RQ2) What percentage of scientific workflows can be reused or not with or without modification?** According to our analysis, 225 (73.3%) workflows among 307 can be reused. On the contrary, 82 (26.7%) workflows are nonreusable due to the challenges discussed in **RQ1(a)**. Figure 12a shows the statistics on the reusability status of the workflows. The ratio of reusable status according to effort level is depicted in Figure 12b. Fortunately, 57.8% of workflows were reusable without any modification. 22.6% of workflows required minor edits and were easy to use. However, 17.8% of workflows needed more changes compared to easily reusable workflows, and only 1.8% of workflows required significant modifications, which were tough to reuse.

### C. Answering RQ3

**(RQ3) How can we overcome the reusability challenges?** A significant number of workflows can be made reusable by conducting some activities on the workflows. We performed several actions to make a workflow reusable through this exploration study. A workflow may need multiple actions based on the issues it contains. Table II shows the action list. We discuss each activity below.

**Identifying proper input dataset:** All operations in a scientific workflow are performed on scientific data. Proper input data selection for a workflow is a crucial task. We identified several workflows where information about the input dataset was missing. In addition, we found a bunch of workflows where input datasets' modules were also unavailable. We identified such datasets by understanding workflow activities. For example, consider a workflow from our sample. The workflow name is *mt analysis 0.01 strand-specific (fastq double)*. Here, there was no input dataset. But from the workflow, it was understandable that the input dataset was a *FASTQ* file. But in many cases, this is not understandable; then, the workflow becomes nonreusable. It is better to annotate the dataset and its source in the workflow step.

TABLE II
ACTIONS TO MAKE A WORKFLOW REUSABLE

| Sl No. | Action List |
|---|---|
| 1 | Identifying proper input dataset |
| 2 | Updating/Upgrading tools |
| 3 | Finding out alternative tools support for obsolete tools |
| 4 | Checking and modifying tools connection |
| 5 | Debugging to find the issue creating tools and connections |
| 6 | Changing step labels |
| 7 | Removing unnecessary portions of workflows |

**Updating/Upgrading tools:** Unfortunately, many workflows were missing the updated/upgraded versions of tools. Workflow publishers should check for any updates/upgrades of the tools used in the workflow.

**Finding out alternative tools support for obsolete tools:** There are many workflows where tools become obsolete in our sample dataset. If a workflow contains an outdated tool, we tried to find another alternative tool to make the workflow reusable. For example, a tool *flagstat* becomes outdated and an enhanced version *Samtools flagstat* is available to support the task of *flagstat*. So we replaced *flagstat* with *Samtools flagstat*.

**Checking and modifying tools connections:** The modules are connected in a workflow, but we noticed quite a few workflows where modules were not connected. Moreover, some connections got broken due to tool up-gradation also. Thus, we added connections among tools wherever necessary.

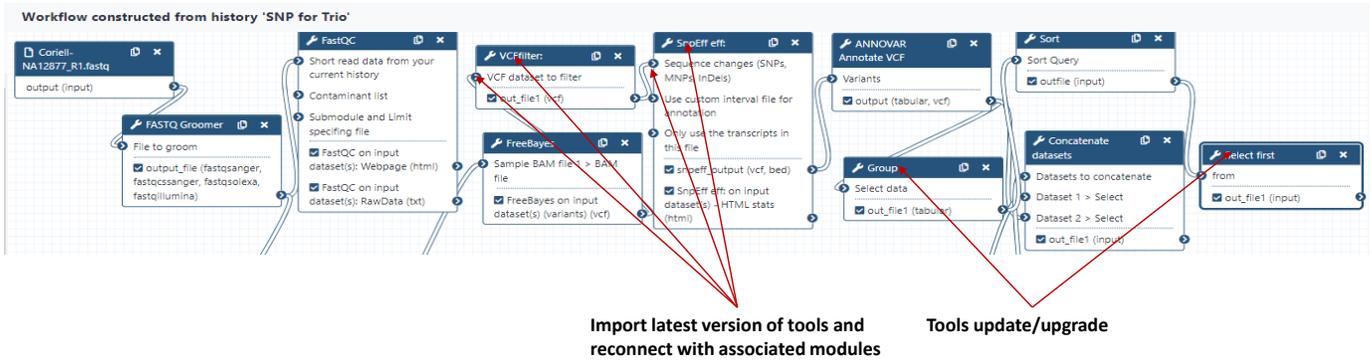

Fig. 11. A segment of a SW (**Reusability status: Difficult to reuse**)

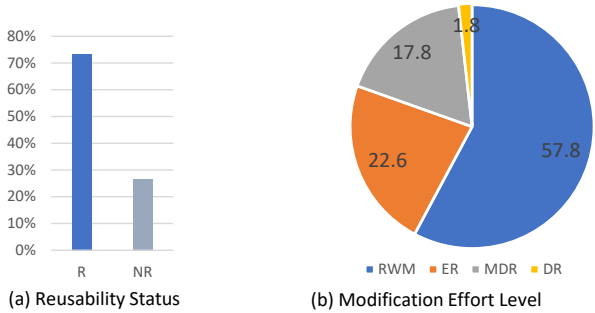

(a) Reusability Status

(b) Modification Effort Level

Fig. 12. Reusability Status (Workflows) & Modification Effort Level (**Reusable Workflows**).

**Debugging to find the issue-creating tools and connections:** We identified several workflows where we needed to perform stepwise debugging to find out issue-creating tools and figured out the issues behind this. Finally, we modified the workflow to make it reusable.

**Changing step labels:** We found a few workflows where several input datasets had the same step labelings for a single workflow. As a result, it generated errors while executing the workflow. Renaming the step labels resolves this issue.

**Removing unnecessary portions of workflows:** SWfMSs support sub-workflow, but we identified some workflows where each workflow contained multiple independent workflows. Keeping the necessary one, we discarded the other workflows.

## VII. KEY FINDINGS AND GUIDELINES

Constructing a SW is a complex and time-consuming task. Reusing can also be very difficult if it is not organized appropriately. Throughout this study, we identified several challenges of reusing a workflow, and below, we provided some key insights to make a workflow reusable without facing any difficulties.

- **Proper description of input dataset and expected outcome:** We identified a large number of workflows where we did not get a clear description of the input dataset. It would be better if the workflow designer could provide a brief annotation about the input dataset, like the sources of data, the nature of the data, and the usage. It is good to have a clear description of the workflow's output.
- **Step-wise annotations:** We found several workflows where the number of modules (tool connections) is too high. Sometimes it is difficult to understand the activities of each module. If step-wise annotations of each module could be provided, it would be easier for others to understand the operations of the workflow and which eventually helps reusability.
- **Update/Upgrade workflow tools on a certain interval:** SWfMSs are evolving extensively. Therefore, tool updates/upgrades become a common phenomenon for SWfMSs to enhance tool capacities, improve performance, and resolve bugs and security flaws. So it would be better if the workflow designers could check their shared workflows on a certain interval to see if any tools require an update/upgrade.
- **Proper parameter values for tools:** Selecting appropriate parameter values is essential to obtain the correct outcome from a workflow. It is recommended to use appropriate parameter values and mention the reasons for choosing them.
- **Check before publishing:** We found workflows in our sample where tools are not connected, multiple irrelevant workflows in a workflow, empty workflows, the purposes mentioned and activities are different, many unnecessary operations, and so on. So before publishing a workflow in the repository, it is recommended to check all the mentioned reasons.

## VIII. THREATS TO VALIDITY

Validity of research is concerned with the question of how the conclusions might be wrong, i.e., the relationship between conclusions and reality [51]. It refers to how sound the research design and methods are. Threats to external validity relate to the generalizability of findings to other populations and settings. We manually analyzed a limited number of scientific workflows related to Galaxy SWfMS, so our results may not generalize to all other workflows from different repositories. Nonetheless, replication of our study with other workflow repositories may prove fruitful. However, we believe that our findings of reusability challenges and actions can be generalized to other workflows from different repositories. But we recommend readers not to over-generalize our findings.

Threats to *internal validity* are influences that can affect the independent variables with respect to causality [52]. The workflow's reusability status (reusable, nonreusable) is threatened by the subjectivity of our classification approach. Thus, we cross-validated our obtained results when a workflow was nonreusable and agreed with 90% similarity.

Threats to *construct validity* are related to the design of the study [53]. We took a sample of 307 workflows; thus, the research outcome may be changed by different samples. To mitigate this issue, we did owner-wise cluster sampling and selected workflows randomly from each cluster. But we avoided taking the same workflows from multiple clusters so that we could explore workflows from various perspectives.

## IX. RELATED WORK

Several research works have been done to design scientific workflow management systems, provenance data management, querying and optimized storing of provenance, and developing querying language. But, the scientific community of workflow management still lacks the reusability of workflows effectively. To simplify creating error-free and effective workflows for scientific data analysis reusability of existing workflows can play a significant role. A few approaches have been taken for reusing existing workflows. Kumar et al. [40] developed a model using a deep learning approach by analyzing workflows available in the European Galaxy server [46] for suggesting tools while constructing a workflow. But they did not consider the issues the workflows may contain. For example, the workflows could be erroneous; there could have incompatible tool connections, obsolete tools and many other challenges. Koop et al. [48] proposed techniques to help scientists construct workflow by automatically suggesting completion based on a database of previously created workflows for the Vistrails system. They developed a tool named *VisComplete*, but they only considered visualization pipelines. They also did not think about the challenges of reusing existing workflows. Junaid et al. [49] proposed a provenance recording and suggestion system for quick and easy creation of error-free workflows. Their approach stored detailed provenance information in the provenance store and provided suggestions based on it to reduce user interaction and automate the design process. But they did not consider the execution of existing workflows. So, while creating workflows, their system can give unexpected tools suggestions. Zhang et al. [4] proposed an approach *Recommend-As-You-Go* for proactively recommending services in a workflow composition process based on the service usage history. But they did not consider the quality aspects of the usage history. Wings [10] can automatically track constraints and rule out invalid designs focusing on the core aspects of the experiments. It generates workflow candidates by searching and validating choices of datasets, parameter values, and software components. It creates multiple variations of workflow using different tools. But, this system ignored the presence of higher-order relationships [47] in tool sequences. Understanding a workflow is difficult due to its complex nature. To facilitate understandability and, therefore, effective reuse of workflows, Garijo et al. [44] identified workflow abstractions from several SWfMSs workflows. But their work did not find the challenges of reusability and actions to overcome them. There is a marked lack in analyzing scientific workflow repositories' reusability challenges. To the best of our knowledge, ours is the first work investigating the reusability challenges and actions to overcome these challenges.

## X. CONCLUSION AND FUTURE WORK

Reusability is essential in scientific workflows because it impacts the efficiency, reproducibility, and scalability of scientific research. The complexity and diversity of workflows across different domains and applications make it difficult to develop a standardized approach to workflow reuse. In this research paper, we manually a statistically significant number of scientific workflows from several repositories and investigated their reusability challenges. Explaining with evidence, we answered three research questions in this work. We classified reusability status into two major categories, namely reusable and nonreusable. We found that 73.29% of workflows can be reused, and 26.71% are nonreusable. We needed to perform several minor or major modifications on 42.22% of reusable workflows, and 57.78% required no change. We also recorded the challenges and described actions to overcome them. In the future, we plan to work with other available workflow repositories, identify their challenges, and resolve them. Then we will develop a workflow recommender system using reusable workflows so that workflow designers can construct efficient and error-free workflows for their purposes.